\documentclass[aps,twocolumn]{revtex4}

\usepackage{amsmath,amssymb}
\usepackage{graphicx}
\usepackage{amsmath,amssymb}
\usepackage{graphicx}
\usepackage{epstopdf}

\newcommand{\be}{\begin{equation}}
\newcommand{\ee}{\end{equation}}
\newcommand{\bea}{\begin{eqnarray}}
\newcommand{\eea}{\end{eqnarray}}
\newcommand{\ba}{\begin{array}}
\newcommand{\ea}{\end{array}}

\def \nn {\nonumber}
\newcommand{\eq}[1]{(\ref{#1})}

\begin{document}
\title{Quantum Secret Sharing with Multi-level Mutually (Un-)Biased Bases}
\author{ I-Ching Yu\footnote{896410029@ntnu.edu.tw},  Feng-Li Lin\footnote{linfengli@phy.ntnu.edu.tw, correspondent author} and Ching-Yu Huang\footnote{896410093@ntnu.edu.tw}}
\affiliation{Department of Physics, National
Taiwan Normal University, Taipei,
116, Taiwan}

\begin{abstract}
We construct general schemes for multi-partite quantum secret sharing using multi-level systems, and find that the consistent conditions for valid measurements can be summarized in two simple algebraic conditions.  The scheme using the very high dimensional mutually unbiased bases can in principle achieve perfect security against intercept-resend attack; and for the scheme using mutually biased bases, it reaches the optimal but non-perfect security at 4-level system.   We also address the security issue against the general attacks in the context of our multi-level schemes. Especially, we propose new protocol to enhance both the efficiency and the security against the entanglement-assisted participant's attack by incorporating quantum-key-distribution and measurement-basis-encrypted schemes so that its security is as robust as quantum-key-distribution.
\end{abstract}


\maketitle


\section{Introduction}
 The security of quantum cryptography is ensured by the non-cloning theorem \cite{noclone} so that the eavesdropping via physical means can always be detected. The schemes for quantum key distribution(QKD) and secret sharing were first proposed in \cite{qkd,Ekert91} and \cite{qss1,qss2}, respectively.  Both of the schemes can be thought as the quantum version of the classical threshold secret sharing $(k,n)$-scheme \cite{ss1,ss2}. The scheme is designed to distribute  valuable information among $n$ participants so that it can be reconstructed only if $k(\le n)$ of them collaborate \footnote{See also \cite{cgl} for deriving the constraint $k>n/2$ on the existence of threshold schemes.}.

  Although the quantum secret sharing(QSS) scheme is better than the classical one in detecting the error caused by an eavesdropper, it is not perfect. For the common intercept-resend attack, an eavesdropper can get hold of some participants' particles, perform the Bell-state measurement \cite{bellm} and resend back. The probability of detecting such an attack is only 25 percent for the QSS scheme \cite{qss1} using 2-level system. The detecting rate is  quite low if the secret sharing is for some fatal event such as the release of warheads, for which we hope to have the perfect security, i.e., 100 percent detecting rate. Therefore, an important question for QSS is whether one can have a scheme with the perfect security for attacks such as intercept-resend.    Surprisingly, despite many modified QSS schemes inspired by the original works \cite{qss1,qss2} in the past few years, as far as we know, there is no discussion for such an issue, even in principle.

  One straightforward way to increase the detecting rate against the attacks is to use higher dimensional quantum systems to proceed the QSS. Intuitively, increasing the dimensions of the quantum bases will complicate the QSS protocol so that the eavesdropper has more difficulty to obtain correct information without being detected. Of course, we will pay the price for reducing the efficiency because we now use the higher dimensional bases to encode one bit information. This may also complicate the consistent conditions for the valid measurements of the protocol and make the QSS procedure more tedious. Moreover, the complication of the protocol may again pose security issues.  

   In this article we construct the QSS schemes using $d$-level systems and establish a security benchmark as a function of $d$ against the common intercept-resend attack. The results show that in principle the perfect security against such an attack can be achieved by using very high dimensional mutually-unbiased bases(MUBs).  Interestingly, we may wonder if the security or error-detecting rate will always increase by using the higher dimensional system. We will see that this is subtle, and we find a counterexample
by using the mutually-biased bases(MBBs) for QSS, which reaches non-perfect optimal security at 4-level system.  Our multi-level scheme is the generalization of \cite{qss1} for 2-level case. It turns out that the consistent conditions for valid measurements of the higher dimensional protocol is quite simple and natural, and can be summarized in two algebraic conditions.   Moreover, regarding the recent concern on the security of QSS \cite{qin}, we will also address the issue in our multi-level schemes against more general and efficient attacks other than intercept-resend. We find that one can invalidate the entanglement-assited participant's attack devised in \cite{qin} by slightly modifying the protocol proposed in \cite{qss1}. 

   The paper is organized as follows: In the next section we will construct the QSS schemes by using the MUBs and MBBs, respectively. In section III we establish a security benchmark against the intercept-resend attack. In section IV we consider the security of our schemes against more general attacks. Especially, we give a proof of security against the attack by an outsider Eve with entangled probe. However, we comment that the original scheme \cite{qss1} is vulnerable to the the entanglement-assited participant's attack devised in \cite{qin}. Finally we conclude our paper in section V by proposing an $100$ percent efficient scheme against the entanglement-assited participant's attack by combining the quantum-key-distribution and measurement-basis-encrypted method.   
   
\section{Quantum secret sharing with mutually unbiased bases}
We first consider the $(2,3)$-scheme for QSS using the multi-level MUBs, and later generalize it to the multi-partite cases. The $(2,3)$-scheme is for Alice to distribute the secret key to both Bob and Charlie, and proceed the QSS protocol via
the local operations and classical communication(LOCC).

We start with the $d$-level GHZ state \cite{ghz}, which is shared among Alice, Bob and Charlie,
\be\label{GHZ}
|GHZ_3\rangle = {1\over \sqrt{d}} \sum^{d-1}_{j=0}|jjj\rangle
\ee
each holds one particle in it. The GHZ state is a maximally entangled state, and is used for QSS such that the measurement outcomes of Alice, Bob and Charlie for their own particles are correlated. Once Bob and Charlie combine their outcomes of measurements, they can know Alice's.  

  A typical QSS protocol  runs as following \cite{qss1,qss2}. (i) Alice prepares the GHZ state, and then distributes the corresponding particles to Bob and Charlie, respectively. (ii) Alice, Bob and Charlie perform the local measurements on their own particles by randomly choosing the measurement bases. (iii) Bob and Charlie then announce their measurement bases publicly to Alice, but not their outcomes. (iv) Alice then determines if the measurement bases satisfy the consistent conditions encoded by the GHZ state, which can be summarized in a lookup table. If so, they keep the outcomes as the useful key and examine further if there is eavesdropping. If there is, then just discard the results. (v) Repeat the above procedure to collect enough outcomes for the secret information. When necessary, Bob and Charlie can collaborate to reproduce Alice's information.

    We will adopt the above protocol for the QSS using $d$-level GHZ state \eq{GHZ}. However, in the end we will modify this protocol to enhance both the security and efficiency of QSS. The modification will incorporate both QKD and measurement-basis-encrypted scheme \cite{pan}.
    
 As noted in \cite{ivanovic,woottersfields}, it is possible to find $d+1$ MUBs in $d$ dimensions only if $d$ is (any power of) an odd prime. Besides the canonical basis $\{|j\rangle, j=0,...,d-1\}$, the explicit forms of the remaining $d$ sets of MUBs are
\be
|P_p\rangle={1\over \sqrt{d}}\sum_{j=0}^{d-1} e^{i\phi(Pj^2+p j)}|j\rangle, \quad \phi={2\pi\over d}
\ee
where $P$ (runs from $1$ to $d$) denotes the basis and $p$ (runs from $0$ to $d-1$) labels the vector in a given orthonormal basis \footnote{The $d=3$ MUBs has been used in \cite{3level} for quantum key distribution.}. They are mutually unbiased because the overlap is
\be\label{muboverlap}
|\langle P_p|P'_{p'}\rangle|={1\over \sqrt{d}}\quad \mbox{for} \quad {P\ne P'},
\ee
which follows from the Gauss sums of number theory valid for odd prime $d$.

  From \eq{muboverlap} we can derive the consistent conditions for a valid measurement.  To arrive that, let us assume that Bob and Charlie hold the states $|B_b\rangle={1\over \sqrt{d}}\sum_{j=0}^{d-1} e^{i\phi(Bj^2+b j)}|j\rangle$ and $|C_c\rangle={1\over \sqrt{d}}\sum_{j=0}^{d-1} e^{i\phi(Cj^2+c j)}|j\rangle$, respectively. Then, take the inner product between the GHZ state and the 2-particle state $|B_b\rangle|C_c\rangle$, we obtain
\be\nn
(\langle B_b|\langle C_c|)|GHZ_3\rangle={1\over d\sqrt{d}}\sum_{j=0}^{d-1}e^{-i\phi((B+C)j^2+(b+c)j)}|j\rangle
\ee
which after normalization should match with Alice's state $|A_a\rangle={1\over \sqrt{d}}\sum_{j=0}^{d-1} e^{i\phi(Aj^2+a j)}|j\rangle$ in order to make a valid measurement for secret sharing.  This then implies the following consistent conditions:
\bea\label{mubc1}
&& A+B+C=0 \;(\mbox{mod}\; d),
\\\label{mubc2}
&& a+b+c=0 \;(\mbox{mod}\; d).
\eea

According to these, we can write down a $d^2 \times d^2$ lookup table for the use of QSS protocol using the $d$-level system. This is the straightforward generalization for the $d=2$ case.   In QSS protocol\cite{qss1,qss2}, one first checks if condition \eq{mubc1} satisfies or not by LOCC. If yes, it is a valid measurement and \eq{mubc2} follows; otherwise, the measurement will be discarded.  Importantly, the conciseness of conditions \eq{mubc1} and \eq{mubc2} helps to simplify the practical en/de-coding procedures in QSS. On the other hand, condition \eq{mubc1} implies that the efficiency is $1/d$ since only one out of $d$ cases makes a valid measurement, and condition \eq{mubc2} can be used to detect eavesdropping for valid measurements.

{\it Quantum secret sharing with mutually biased bases.---}
   The above generalizes the QSS scheme of \cite{qss1,qss2} using MUBs. Now we look for the scheme using MBBs which has not been discussed before in literatures.

   Our construction of the $d$-level MBBs for QSS is as follows. Start with the Fourier transform of the canonical basis
\be\nn
|u_k\rangle_F={1\over \sqrt{d}}\sum_{j=0}^{d-1}e^{i k j \phi}|j\rangle, \quad k=0,..,d-1, \quad \phi={2\pi \over d}.
\ee
We then introduce the following $d$ MBBs
\be\label{biased}
|P_p\rangle=|u_p\rangle_F+{1\over\sqrt{d}}(e^{iP\phi}-1)|0\rangle
\ee
for $P=0,..,d-1,p=0,..,d-1$. Note that $|0_j\rangle=|u_j\rangle_F$. For simplicity, here and hereafter we use the same notation as for MUBs.

  We now show that these bases will form a consistent lookup table for QSS. Note that the overlap
\be\label{overlappp}
\langle P_p|P'_{p'}\rangle =\delta_{pp'}+{1\over d}[e^{i(P'-P)\phi}-1].
\ee
Thus, each basis $\{|P_p\rangle, p=0,..,d-1\}$ is orthonormal and complete, and the overlap $|\langle P_p|P'_{p'}\rangle|$ between different bases will depend on $P'-P$ and so is called biased except for $d=3$ case, which is the same as $d=3$ MUBs'.

  Similar to the case for MUBs, if Bob and Charlie hold the states $|B_b\rangle=|u_b\rangle_F+{1\over\sqrt{d}}(e^{i B\phi}-1)|0\rangle$ and $|C_c\rangle=|u_c\rangle_F+{1\over\sqrt{d}}(e^{i C\phi}-1)|0\rangle$ respectively, we should require the state $(\langle B_b|\langle C_c|)|GHZ_3\rangle$ to match Alice's state $|A_a\rangle=|u_a\rangle_F+{1\over\sqrt{d}}(e^{i A\phi}-1)|0\rangle$ for a valid measurement. This then yields the same consistent conditions \eq{mubc1} and \eq{mubc2} for a valid measurement as for MUBs.

   It is straightforward to generalize the above tri-partite scheme to the multi-partite one.
We just start with the $n$-partite GHZ state
\be\label{dGHZ}
|GHZ_n\rangle={1\over \sqrt{d}}\sum_{j=0}^{d-1}|j j j... j\rangle_{123...n}.
\ee
Each party measures her/his own particle and obtains the outcome in one of the $d$ bases, say $\{|P_p\rangle\}$. To have a consistent
lookup table for $n$-partite case, we should require (up to some normalization factor)
\be\nn
|A_a\rangle= (\langle B_b| \langle C_c| \langle D_d|\cdots \langle \Omega_{\omega}|)|GHZ_n\rangle
\ee
which yields the following consistent conditions
\bea\label{dtab3}
&& A+B+C+...+\Omega=0 \;(\mbox{mod}\; d),
\\\label{dtab4}
&& a+b+c+...+\omega=0 \;(\mbox{mod}\; d).
\eea
These are straightforward generalization of \eq{mubc1} and \eq{mubc2}, respectively. Note that because of \eq{muboverlap} or \eq{overlappp}, condition \eq{dtab3} implies \eq{dtab4} but not vice versa.

\section{Detecting the intercept-resend attack}
  We like to give a benchmark formula for the detecting rate against the very common intercept-resend attack as a function of dimension $d$. The attack goes as follows for tri-partite case:  the dishonest Charlie*
gets hold of Bob's particle and performs the general Bell-state measurement on his two-particle state, then resends one particle to Bob. Since Charlie* does not know Alice's measurement basis, he may use the wrong base for his Bell-state measurement but still has some probability to get the right result. On the other hand, if Charlie* happens to use the right base, he will then know Bob's measurement outcome and then Alice's after LOCC without making detectable error.

   The detecting rate against the attack can be derived as follows. Let us assume that Alice's measurement outcome is $|A_a\rangle$, However, Charlie* thinks Alice was using the base $\{|A'_{a'}\rangle\}$ and expands his two-particle state in such a base, i.e.,
$\langle A_a|GHZ_3\rangle= \sum_{a'=0}^{d-1} \langle A_a|A'_{a'}\rangle \langle A'_{a'}|GHZ_3\rangle$.
A detectable error occurs if the condition \eq{mubc1} holds but \eq{mubc2} is violated, and its rate is
$1-|\langle A'_{a'}|A_a\rangle|^2$.
Then, the average detecting rate over the configurations satisfying \eq{mubc1} but not \eq{mubc2} is
\be\label{pe1}
P_E:=\sum_{A-A'=1}^{d-1} {1\over d} \sum_{a'=0}^{d-1} |\langle
A_a|A'_{a'}\rangle|^2 (1-|\langle A'_{a'}|A_a\rangle|^2).
\ee

For the scheme using MUBs, by \eq{muboverlap} the detecting rate \eq{pe1}
is
\be\label{pe3}
P_{E,MUBs}(d)=(\frac{d-1}{d})^2 .
\ee
It is a monotonically increasing function of $d$ so that higher dimensional system is more secure.
Especially, it approaches unity as $d$ goes to infinity and implies perfect security, in principle.

For the cases using MBBs, by \eq{overlappp} we find that
\be\label{pe2}
P_{E,MBBs}(d)={4d^2-10d+6 \over d^3}.
\ee
In contrast to MUBs' case, it is not a monotonic function of $d$ because of the weighted overlap between bases. Instead, it reaches the
maximal at $d=4$ with $P_{E,MBBs}(d=4)={15\over 32}$, and then decreases to zero monotonically for $d>4$. Moreover, $P_{E,MUBs}(d=3)=P_{E,MBBs}(d=3)={4\over9}$ as expected because our MUBs and MBBs are the same for $d=3$. Recall that the detecting rate for the 2-level scheme of \cite{qss1} is only $1/4$, so even the lower $d(>2)$ schemes using MBBs have higher security than the 2-level case. Since the MUBs' scheme is only available for odd prime $d$, the $d=4$ MBBs' case can be considered as the optimal scheme for a compromise between the degree of security and the efficiency. Practically, one can physically realize the $d=4$ system by combining two 2-level systems to carry out the optimal MBBs' scheme; for example, the para- and ortho-helium spectra can be seen as a $d=4$ system by appropriately adjusting the external magnetic/electric fields.

 Estimating the detecting rate in the multi-partite system is more complicated,  we will not discuss the details here.
 However, the more persons share the
entangled key, the more difficult it is for the eavesdropper to
collect all the other's particles and the more secure the scheme is.

\section{Security against general eavesdroppers}
Enforcing the security of the cryptography is state-of-the-art, so is its attack strategy. After establishing a security benchmark against the common intercept-resend attack, we like to address the issue for more general attacks which could be more efficient than expected for
an eavesdropper (called Eve) using the ancilla probe.

 First, we consider the case that Eve is not the member sharing the secret via GHZ state. Then, the QSS scheme is secure provided that GHZ state is the only state satisfying the consistent conditions \eq{dtab3} and \eq{dtab4}.  Otherwise, there will be a set of fake key states $\{|FK\rangle\}$ other than the GHZ  satisfying  \eq{dtab3} and \eq{dtab4}, and Eve can use the ancilla states $\{|E\rangle_{FK}\}$ and $|E\rangle_{GHZ}$ to form the entangled state
\be\label{general eve}
|\Psi\rangle=|GHZ_n\rangle|E\rangle_{GHZ}+\sum_{\{|FK\rangle\}} |FK\rangle|E\rangle_{FK}.
\ee
She can then extract the encoded secret information in the GHZ state by performing the general Bell-state measurement without making detectable errors. We now show that  GHZ state is indeed the unique one satisfying \eq{dtab3} and \eq{dtab4}.

   The proof constructed for the 2-level scheme is given in \cite{qss1} by showing that all the states orthogonal to the GHZ state do not satisfy the consistent conditions \eq{dtab3} and \eq{dtab4}.  This procedure will be far more involved for the multi-level case.  Instead, we directly show
\be\label{secure1}
|\langle \Lambda|\Phi\rangle|< |\langle \Lambda|GHZ_n\rangle|
\ee
for any state $|\Phi\rangle$ belonging to the vector space $V^{\perp}_{GHZ}$ which is orthogonal to GHZ state, and the conditional states $|\Lambda\rangle$'s representing the consistent conditions \eq{dtab3} and \eq{dtab4}, i.e.,
\be
|\Lambda\rangle=|A_a\rangle|B_b\rangle|C_c\rangle...|\Omega_\omega\rangle, \nn
\ee
with the states' labels satisfying \eq{dtab3} and \eq{dtab4}. This implies that none of the states in $V^{\perp}_{GHZ}$ will satisfy all the $d^{n-1}$ conditions given by \eq{dtab3} and \eq{dtab4}. The state $|\Psi\rangle$ in \eq{general eve} will then reduce to the product of GHZ  and ancilla states so that Eve cannot obtain useful information through entanglement without making detectable errors.

  We start the proof by constructing the basis vectors for $V^{\perp}_{GHZ}$ in terms of the canonical basis $\{|ijk\cdots\rangle\}$ via Gram-Schmidt orthonormalization process. Then, there arrive two kinds of basis states: (1) there are $d^n-d$ unit vectors in the canonical basis which do not belong to the subset made of GHZ state, i.e.,
\be
|\Phi\rangle_{\perp,1}:=|ijk\cdots\rangle \ne |\ell\ell\ell\cdots\rangle \nn
\ee
with  $i,j,k,\ell \cdots =0,1,2,\dots,d-1$;
(2) there are the other $d-1$ basis vectors taking the following form
\be
|\Phi\rangle_{\perp,2}=\sum_{j=0}^{d-1} c_j|jjj\cdots\rangle, \nn
\ee
and the orthogonality to GHZ state requires $\sum_{j=0}^{d-1} c_j=0$
besides the normalization condition $\sum_{j=0}^{d-1} |c_j|^2=1$.

   Before we check if states $|\Phi\rangle_{\perp,1}$ and $|\Phi\rangle_{\perp,2}$ satisfy \eq{secure1}, we note that  
\be \label{concg1}  
\langle \Lambda|jjj\cdots\rangle=d^{-{n\over 2}} 
\ee
for any conditional state $|\Lambda\rangle$ so that
\be\label{concg2}
\langle \Lambda|GHZ_n\rangle=d^{1-{n\over 2}}.
\ee
These two imply that while checking \eq{secure1} we can treat all the $d^{n-1}$ conditional states equally, it then helps to simplify the task.  Condition \eq{concg1} then yields
$\langle \Lambda|\Phi\rangle_{\perp,2}=d^{-{n\over 2}}\sum_{j=0}^{d-1} c_j=0$ for any $|\Lambda\rangle$ so that the second type of basis vectors are orthogonal to all the conditional states and thus are excluded from the set $\{|FK\rangle\}$ in \eq{general eve}.  On the other hand, from \eq{concg2} for the first kind of basis vectors we have $
|\langle \Lambda|\Phi\rangle_{\perp,1}|=d^{-{n\over 2}}<\langle \Lambda|GHZ_n\rangle$
for any conditional state $|\Lambda\rangle$ so that they are also excluded from the set $\{|FK\rangle\}$ in \eq{general eve}. This then completes our proof. 

  However, the uniqueness of GHZ state does not guarantee QSS' security if Charlie* is dishonest with the help of an ancilla Eve to entangle with Bob's particle. This is the entanglement-assisted participant's attack. In \cite{qin}, an explicit attacking scheme via manipulating GHZ state with ancilla was devised so that $AB$'s and $CE$'s states  are maximally entangled \footnote{In \cite{qin} the authors consider $d=2$ case, it is straightforward to generalize to d-level cases for both MUBs and MBBs by using the d-level Hadamard and CNOT gates. Also note that the form of \eq{abce} is maximally entangled in Schmidt's decomposition between two parties $AB$ and $CE$.} 
\be\label{abce}
|\Psi\rangle_{ABCE}={1\over d} \sum_{a,b=0}^{d-1} |\bar{A}_a\rangle |\bar{B}_b\rangle \otimes |\psi^{(\bar{A}\bar{B})}_{ab}\rangle_{CE}
\ee
where the bar quantity means its value is chosen and fixed, and $\{|\psi^{(\bar{A}\bar{B})}_{ab}\rangle_{CE}\}$ is an orthonormal complete set for Alice's and Bob's chosen basis $\bar{A}$ and $\bar{B}$, respectively. After Alice and Bob measure their particles in bases $\bar{A}$ and $\bar{B}$ with the outcomes $a=\bar{a}$ and $b=\bar{b}$ respectively, then the state \eq{abce} collapses to $|\psi^{(\bar{A}\bar{B})}_{\bar{a}\bar{b}}\rangle_{CE}$. Since  $\{|\psi^{(\bar{A}\bar{B})}_{ab}\rangle_{CE}\}$ is orthonormal and Charile* knows Alice's and Bob's measurement basis, he can perform local unitary transformations to extract the    
$\bar{a}$ and $\bar{b}$ from $|\psi^{(\bar{A}\bar{B})}_{\bar{a}\bar{b}}\rangle_{CE}$ without making detectable error as shown in \cite{qin}. In a sense, the multi-level QSS scheme using protocol of \cite{qss1} is highly insecure.  

\section{An efficient scheme against entanglement-assisted participant's attack}
   We now propose a modified protocol to remedy the above security loop-hole. Moreover, it will enhance the efficiency of QSS from $1/d$ to 100 percent. The modification is two-fold. One is to adopt the measurement-basis-encrypted efficient QSS scheme proposed in \cite{pan} as follows: Instead for the participants to announce their measurement basis in order to verify if they satisfy \eq{dtab3} for the valid measurements, they will use their measurement outcomes as the measurement basis for the next run. As long as the first run is a valid measurement, then all the subsequent runs will be automatically the valid measurements as seen from \eq{dtab3} and \eq{dtab4}. This yields 100 percent efficiency. Moreover, since Charile* does not know about others' chosen bases and thus cannot take advantage of the entangled state \eq{abce}, he has no way to extract other's measurement outcomes from such a state $|\psi^{(\bar{A}\bar{B})}_{ab}\rangle_{CE}$. By guess, he has only $1/d$ chance to do it right. The remaining modification is to ensure the first run can be a valid measurement without announcing the measurement basis. This can be done by using multi-level QKD for Alice to distribute a valid set of measurement basis to each participant separately. The eavesdropper can of course attack QKD, too. However, the QKD security is more robust than QSS's and has been studied extensively, e.g. see \cite{gisin}.  An alternative against the attack of \cite{qin} is considered in \cite{chi} for multi-level QSS recently. 
   
   In this paper, we have generalized the QSS scheme for qubits to the multi-level cases with both MUBs and MBBs. We also discuss the security issues for general attacks. Finally we propose an efficient and secure protocol which could be relevant to the physical realization of QSS.
   
\bigskip
   
{\bf Acknowledgement:}  We would like to thank Ming-Che Chang, Li-Yi Hsu and I-Ming Tsai for valuable comments. This work was supported by Taiwan's NSC grant 96-2112-M-003-014.



\begin{thebibliography}{99}

\bibitem{noclone}
W.~K.~Wootters and W.~H.~Zurek, Nature {\bf 299}, 802 (1982).
D.~Dieks, Physics Letters A, {\bf 92}271 (1982).

\bibitem{ss1}
 G.~R.~Blakley, in Proceedings of the American
Federation of Information Processing 1979 National Computer
Conference (American Federation of Information Processing,
Arlington, VA, 1979), pp.313-317.

\bibitem{ss2}
A.~Shamir, Commun. ACM {\bf22},
612 (1979).

\bibitem{qkd}
C.~Bennett and G.~Brassard in
\emph{Proceedings of IEEE International Conference on
Computers, Systems, and Signal Processing} (Bangalore,
India 1984) p. 175.

\bibitem{Ekert91}
A.~K.~Ekert, Phys.\ Rev.\ Lett.\ {\bf 67}, 661(1991).

\bibitem{qss1}
M.~Hillery, V.~B\v{u}zek, and A. Berthiaume, Phys.
Rev. A {\bf59}, 1829(1999).

\bibitem{qss2}
A.~Karlsson, M.~Koashi, and N.~Imoto, Phys. Rev. A
{\bf59}, 162 (1999).

\bibitem{cgl} R.~Cleve, D.~Gottesman, and H.-K. Lo,
Phys.\ Rev.\ Lett.\ {\bf 83} 648 (1999).


\bibitem{ghz}
D.~ Greenberger, M.~Horne, and A.~Zeilinger in \emph{Bell's Theorem, Quantum Theory, and
Conceptions of the Universe} (Kluwer Academic, Dordrecht, 1989).

\bibitem{ivanovic}
I.~D.~Ivanovic, J.\ Phys.\ A {\bf 14}, 3241 (1981).

\bibitem{woottersfields}
W.~K.~Wootters and B.~D.~Fields, Annals of Physics, {\bf 191}, 363 (1989).


\bibitem{3level}
H.~Bechmann-Pasquinucci and A.~Peres, Phys.\ Rev.\ Lett.\ {\bf85}, 3313 (2000).

\bibitem{bellm}
M.~Michler, K.~Mattle, H.~Weinfurter, and A.~Zeilinger, Phys. Rev. A {\bf 53}, R1209(1996).

\bibitem{qin}
S.-J.~Qin, F.~Gao, Q.-Y.~Wen, F.-C.~Zhu, Phys. Rev. A {\bf76}, 062324(2007).

\bibitem{pan} L.~Xiao, G.-L.~Long, F.-G.~Deng, J.-W.~Pan,  Phys. Rev. A {\bf69}, 052307 (2004).

\bibitem{gisin} N.~J.~Cerf, M.~Bourennane, A.~Karlsson, and N.~Gisin, Phys.\ Rev.\ Lett.\ {\bf 88}, 127902 (2002).

\bibitem{chi}
D.-P.~Chi, J.-W.~Choi, J.-S.~Kim, T.-W.~Kim, S.-J~Lee, arXiv:0801.0177.

\end{thebibliography}
 \end{document}